\documentclass{iau}
\usepackage{graphicx}

\title[The Binggeli effect] 
{The Binggeli effect}

\author[Biernacka M. et al.]   
{Biernacka M.$^{1}$,
Panko E.$^{2}$, God{\l}owski W.$^{3}$, Bajan K.$^{4}$, Flin P.$^{1}$}
\affiliation{$^1$Institute of Physics,  Jan Kochanowski University, Kielce, Poland \\
email: {\tt bmonika@ujk.edu.pl} \\[\affilskip]
$^2$Kalinenkov Observatory, Nikolaev National University, Nikolaev, Ukraine \\[\affilskip]
$^3$Institute of Physics, Opole University, Opole, Poland \\[\affilskip]
$^4$Mt. Suhora Observatory, Cracow Pedagogical University,
Krakow, Poland \\[\affilskip]

}

\pubyear{2008}
\volume{xxx}  
\pagerange{119--126}
\jname{Title of your IAU Symposium}
\editors{A.C. Editor, B.D. Editor \& C.E. Editor, eds.}
\begin{document}

\maketitle

\begin{abstract}
We found the alignement of elongated clusters of BM type I and III (the excess of small values of the $\Delta\theta$ angles is observed), having range till about $60 h^{-1}Mpc$. The first one is probably connected with the origin of supergiant galaxy, while the second one with environmental effects in clusters, originated on the long filament or plane.
\end{abstract}

\firstsection 
\section{Introduction}

\cite[Binggeli (1982)]{Binggeli1982} was the first who found that galaxy clusters tend to be aligned pointing each other. Later on the existence of this effect was discussed by several authors and usually the significant alignment was reported. The distance between clusters for which the effect was detected changed from $10h^{-1} Mpc$ till $150h^{-1} Mpc$ and the strength of the effect diminished with distance (\cite[Struble \& Peebles (1985)]{Struble}, \cite[Flin (1987)]{Flin}, \cite[Rhee \& Katgert (1987)]{Rhee}, \cite[Ulmer et al. (1989)]{Ulmer}, \cite[Plionis (1994)]{Plionis}, \cite[Chambers et al. (2002)]{Chambers}). These investigation employed both optical and X-ray data, as well as clusters assigned or not to superclusters. Nowadays it is accepted that the effect is real, not due to selection effects, and it distance scale is between $(10 - 60) h^{-1}Mpc$. Better understanding of physical processes leading to cluster alignment revealed numerical simulations. This was done in the framework of cold dark matter (CDM) model regarded now as being correct for large scale structure fomation (\cite[Onuora \& Thomas (2000)]{Onuora}) using large - scale simulation found that in LCDM cosmological model for distance up to $30 h^{-1}Mpc$, while in tCDM models the range of effects is twice smaller. In SCDM and OCDM models, where smaller scale simuations were performed some alignment effect could be noted.
\section{Observational data}
The catalogue of 6188 galaxy clusters (hereafter PF,  \cite[Panko \& Flin (2006)]{Panko})  served as observational basis of our studies. It is prepared applying Voronoy  tessellation technique (\cite[Ramella et al. 1999, 2001]{Ramella}) to Muenster Red Sky Survey (MRSS, \cite[Ungruhe et al. (2003)]{Ungruhe}). The PF  catalogue is statistically complete till  $r_F =18.3^m$. Our cluster  contains at  least 10 galaxies in the brightnest  range $m_3$, $m_3 + 3^m$ ($m_3$ is the magnitude of the third brightest galaxy). The distance to clusters were determined using the  dependence between the tenth brightest galaxy $m_{10}$ and redshift (\cite[Panko et al. (2009)]{Panko1}). Clusters morphological types for 1056 PF objects were taken from ACO (\cite[Abell et al. (1989)]{Abell}).
\section{Method and results}
Method used to determine galaxy clusters shape (ellipticity and position angle) was covariance ellipse method \cite[(Carter \& Metcalfe (1980))]{Carter}. The method is based on the first five moments of the observed distribution of galaxy coordinates $x_i$, $y_i$. The existence of the Binggeli effect was checked studing the angle $\Delta\theta$ between the direction toward neighbouring clusters and the cluster position angle. The coordinate distances between clusters were calculated assuming $h=0.75$ , $q_o=0.5$.
\begin{figure}[h]
\begin{center}
 \includegraphics[width=1.2in]{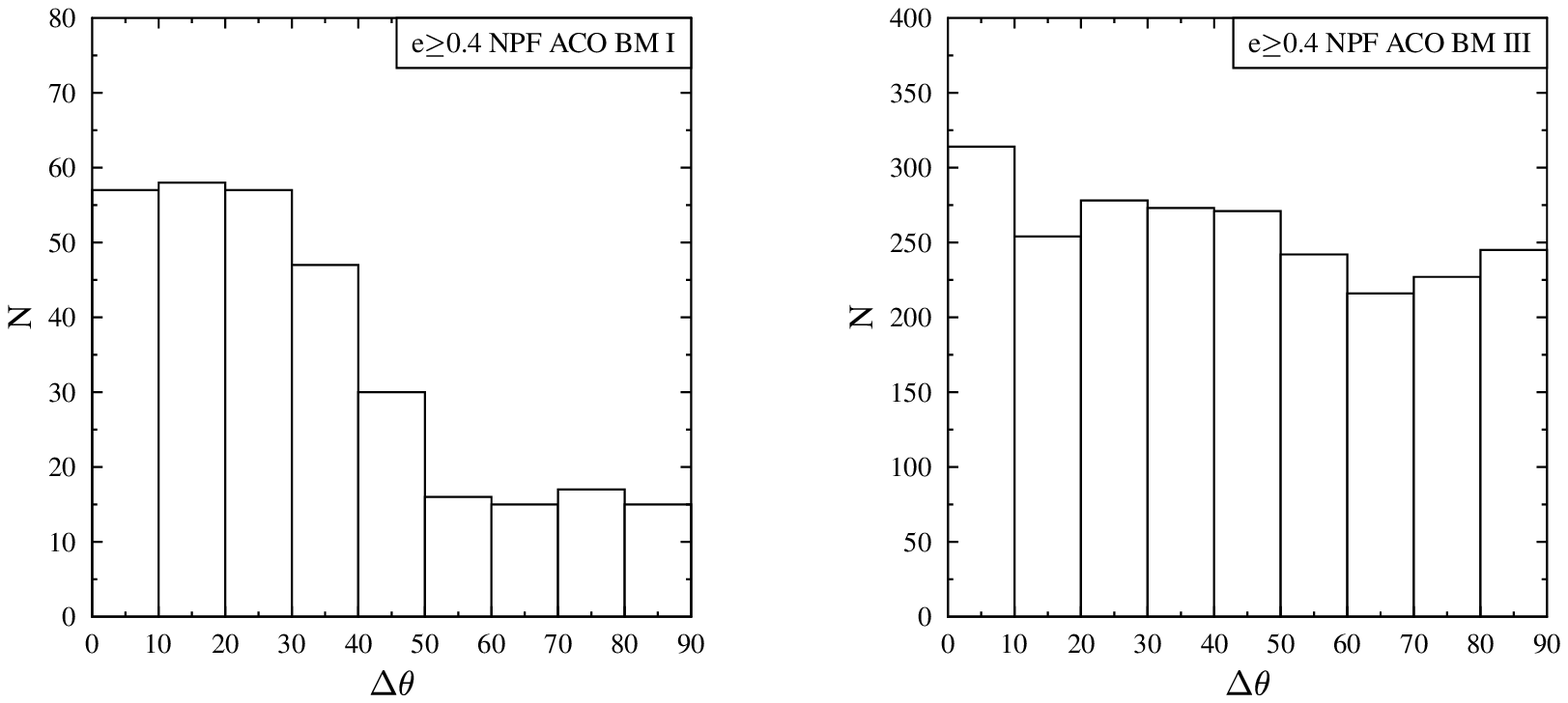} 
 \caption{Distribution of the angle $\Delta\theta$ for BM type I and III with galaxy clusters ellipticities $e>0.4$.}
   \label{fig1}
\end{center}
\end{figure}
\section{Conclusions}
\begin{itemize}
\item We found the alignment of elongated clusters of BM type I and III (the excess of small values of the $\Delta\theta$ angles is observed), having range till about $45 h^{-1} Mpc$. The BMI was expected due to special role of CD galaxies during formation. The existence of the Binggeli effect in BM III is surprising. It can be due to incorporation of RS types L and F in which alignment is frequently observed.
\item The effect is observed only in the sample containing elongated clusters. Cluster ellipticity is not well defined parameter, which was noted by several investigators. The problem with cluster ellipticity determination is not only the problem of the applied method.
\item The alignment in BMI is probably connected with the origin of supergiant galaxy, while in the case of BMIII with environmental effects in clusters originated on the one long filament or plane.
\item It is quite possible that the origin of galaxy clusters was not a unique, homogeneous process and various factors influenced this process.
\end{itemize}

\end{document}